\documentclass[referee]{aa} 

\usepackage{graphicx}
\usepackage{txfonts}
\usepackage{color}

\usepackage{natbib}
\bibpunct{(}{)}{;}{a}{}{,}

\def \period {$3.4530\,\pm\,0.0014$ days }

\begin{document}

   \title{Photometric Analysis of the Optical Counterpart of the Black Hole HMXB M33 X-7}

   \titlerunning{Photometric Analysis of M33 X-7 Optical Component}

   \author{A. Shporer \inst{1}
           \and
           J. Hartman\inst{2}
           \and
           T. Mazeh \inst{1}
           \and
           W. Pietsch \inst{3}
          }
 
   \offprints{shporer@wise.tau.ac.il}

   \institute{Wise Observatory, Tel Aviv University, Tel Aviv, Israel 69978
              \and
              Harvard-Smithsonian Center for Astrophysics, 60 Garden St., Cambridge, MA 02138, USA        
              \and
	      Max-Planck-Institut fur extraterrestrische Physik, 85741 Garching, Germany      
}


  \abstract
   {}
   {Study the high-mass X-ray binary X-7 in M33 using broad-band 
    optical data.}
{We used recently published CFHT $r'$ and $i'$ data for variable stars
   in M33 to extract the light curve of the optical counterpart of
   X-7. We combined these data with DIRECT $B$ and $V$ measurements in
   order to search for an independent optical modulation with the
   X-ray periodicity.  The periodic modulation is modelled with the
   ellipsoidal effect. We used $UBVRr'i'$ magnitudes of
   the system to constrain the temperature and radius of the optical
   component.} 
{ The optical data revealed a periodicity of \period, which is
   consistent with the known X-ray period. Double modulation, which we
   attributed to ellipsoidal modulation, is clearly seen in four
   different optical bands.  The absolute magnitude in six optical
   bands is most consistent with a stellar counterpart with $ 33000 <
   T_{eff} <~\!\!47000 $\,K and $15 < R < 20\,R_{\sun}$. We modelled
   the optical periodic modulation and derived the masses of the two
   components as a function of the orbital inclination and the radius
   of the stellar component. The resulting mass range for the compact
   object is $1.3 < M < 23\,M_{\sun}$.  }
{The system is probably a black hole HMXB, similar to Cyg X-1, LMC X-1
and LMC X-3.}
\maketitle
\section{Introduction}
\label{intro}

M33 X-7 (hereafter X-7) is a well known bright variable X-ray source
in the nearby M33 galaxy \citep{Long81}. Its variability was identified 
as periodic by \cite{Peres89}, who derived a period of 1.7857 days and 
were the first to suggest its XRB nature.
Using data from {\it Einstein}, ROSAT and ASCA, \cite{Larson97} have
suggested that the actual period is $3.4531$ days. This was confirmed
by \cite{Dubus99}, who reported a period of $3.4535 \pm 0.0005$ days.

\cite{Pietsch04b} have analyzed X-ray data from XMM-Newton and {\it
Chandra} and derived an improved period of $3.45376 \pm 0.00021$
days. They also used optical $B$ and $V$ data to suggest the optical
companion is of spectral type between B0 I and O7 I, with masses of
25--35 $M_{\sun}$, making the compact binary component a black
hole. If this is true, X-7 is the first eclipsing black hole high mass
X-ray binary (HMXB).

In the course of the {\it Chandra} ACIS-I survey of M33
\citep[ChASeM33,][]{Sasaki05a}, the X-ray eclipse ingress and egress
of X-7 were observed for the first time \citep{Sasaki05b}.
\cite{Pietsch06} combined {\it Einstein}, ROSAT and XMM-Newton data,
and derived an improved period, of $3.453014 \pm 0.000020$ days and a
possible period decay rate of $\dot{P}/P = -4 \times
10^{-6}$~yr$^{-1}$.  \cite{Pietsch06} have also used HST WFPC2 images
to resolve the OB association HS 13 \citep{Humphreys80} in which X-7
resides, and have identified the optical counterpart to be an O6 III
star whose mass is at least 20 $M_{\sun}$. Based on the optical
counterpart identification, possible orbital period decay rate, lack
of pulsations, X-ray spectrum and fits to the optical light curves of
\cite{Pietsch04b}, \cite{Pietsch06} suggested that X-7 is a black hole
HMXB.

In the previous studies the periodicity of X-7 was derived from the
X-ray data, and no independent analysis of optical data was
performed. Optical light curves of X-7 were presented in $B$ and $V$
filters by \cite{Pietsch04b}, who folded the optical measurements on
the X-ray period. This paper presents an analysis of optical light
curves of the X-7 counterpart obtained by \cite{Hartman06} in a search
for variable stars in M33, performed at the
Canada-France-Hawaii-Telescope (CFHT) with the Sloan $r'$ and $i'$
filters. \cite{Hartman06} measurements are combined here with the $B$
and $V$ observations of \cite{Pietsch04b} to search independently for
a periodic modulation. The combined data, as presented in
\S~\ref{data} do show clear periodicity with the X-ray
period, as presented in \S~\ref{periodanalysis}. Based on broad-band
$UBVRr'i'$ data from \cite{Pietsch06}, \cite{Massey06} and
\cite{Hartman06} we apply a photometric modelling to derive the stellar
temperature and radius of the optical component in
\S~\ref{photometricmodelling}, and periodic ellipsoidal model in
\S~\ref{lightcurvemodelling}. We briefly discuss our results in
\S~\ref{summary}.


\section{The Optical data}
\label{data}

\begin{figure}
  \centering
  \includegraphics[width = 6cm]{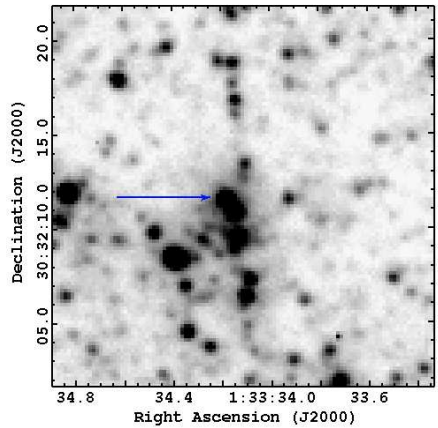}
  \caption{X-7 finding chart, in $i'$, from the CFHT data. Position of the 
    variable object is indicated by an arrow. Image FOV is $20 \times 20$ 
    arcsec. North is up and East is to the left.}
  \label{X-7i}
\end{figure}

The present analysis is based on the optical measurements of X-7 in
four bands, obtained from two sets of observations carried out
$\sim~\!\!\!\!4$ years apart. The $B$ and $V$ light curves of
\citet[][ Table 4]{Pietsch04b} were obtained from a special
photometric analysis of
DIRECT\footnote{$\!\!$http://cfa-www.harvard.edu/$\sim$kstanek/DIRECT/}
\citep{Mochejska01b} images, taken at the Kitt Peak National
Observatory 2.1-m telescope in 1999. The $V$ light curve includes 70
measurements and the $B$ light curve only 30 points. Sloan $r'$ and
$i'$ light curves were taken from the publicly available data of the
M33 CFHT variability
survey\footnote{$\!\!$http://www.astro.livjm.ac.uk/$\sim$dfb/M33/}
\citep[][ object 243718]{Hartman06}. Their data were obtained at the
CFHT 3.6-m telescope atop Mauna Kea on two consecutive observing
seasons in 2003 and 2004. An $i'$ finding chart of X-7 is given in
Fig.~\ref{X-7i}. A comparison with the finding chart of \citet[][
their figure 5]{Pietsch06} shows this object is at the position of X-7
optical counterpart identified by \citet{Pietsch06}. The
identification of the same period in the optical as in the X-rays
(presented hereafter) confirms the identification. The $i'$ light
curve contains 32 measurements, where we ignored two measurements with
uncertainty larger than 0.04 mag. The $r'$ light curve consists of 31
measurements, where we ignored three obvious outliers.

\section{Period analysis}
\label{periodanalysis}

To search for periodicity we have applied a multi-band double-harmonic
fitting \citep{Shporer06} to the four available light curves together,
which included 158 individual measurements. For each trial period we
fitted the entire data set with a double-harmonic function, with
independent zero points for each of the four light curves. The
periodogram value for each trial period was taken as the power (sum of
squared harmonic coefficients) divided by the $\chi^2$ goodness-of-fit
parameter of the fitted light curve.

\begin{figure}
  \centering
  \includegraphics[width = 9cm]{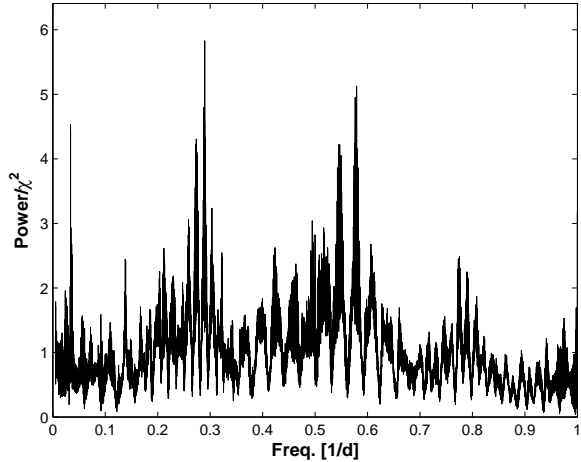}
  \caption{Multi-band double-harmonic periodogram.}
  \label{periodogram}
\end{figure}

Fig.~\ref{periodogram} presents the resulting periodogram. The
strongest peak corresponds to a period of $P_{opt}$ = \period. The
second strongest peak is exactly at the first harmonic of
$P_{opt}$. The third strongest peak is at a frequency corresponding to
the lunar cycle.  The optical period agrees well with the known X-rays
period, $P_X = 3.453014 \pm 0.000020$ days, of \cite{Pietsch06}.

\begin{figure}
  \centering
  \includegraphics[width = 9cm]{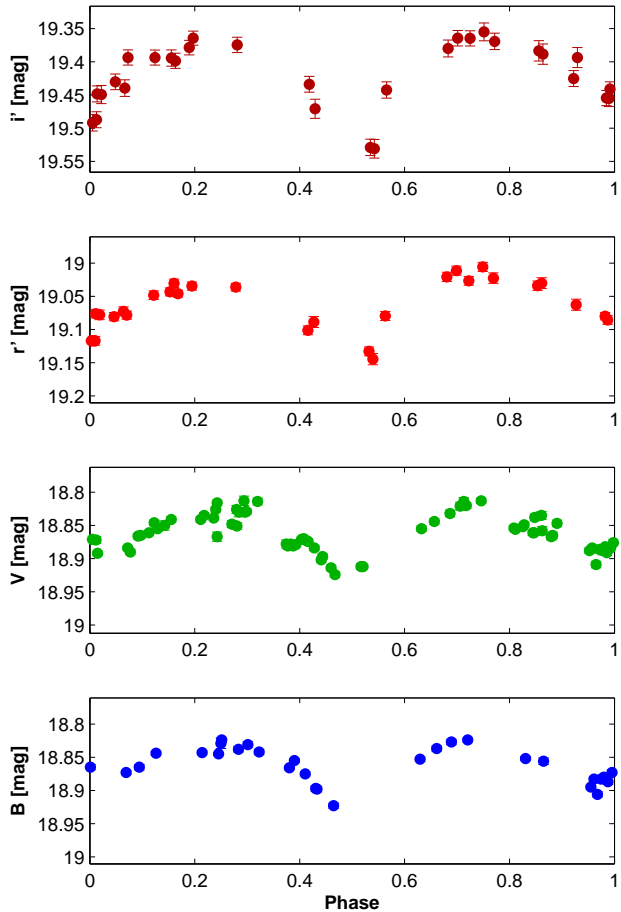}
  \caption{Folded light curves, in magnitude, in (top to bottom) $i'$,
    $r'$, $V$ and $B$.  All light curves are folded on the period
    identified here, of 3.4530 days, and phase zero was taken to be at
    HJD = 2453639.119, which is time of mid-eclipse in the X-rays
    \citep{Pietsch06}.}
  \label{lc}
\end{figure}

Fig.~\ref{lc} presents the four light curves folded on the detected
period, with phase zero taken at HJD = 2453639.119 --- the mid-eclipse
time of \cite{Pietsch06}. The figure shows the $B$ and $V$ data of
\citet{Pietsch04b} for completeness, including seven $V$ measurements
that were not plotted in figure 4 of \citealt{Pietsch04b}, since that
figure brings only $V$ measurements that were obtained close to $B$
measurements. Fig.~\ref{lc} clearly shows ellipsoidal modulation in
all four bands.  For each band independently, we fitted the data with
two harmonics and derived the total amplitude of the modulation and
the phase of minimum. The results are listed in
Table~\ref{harmonicfit}. For the $i'$ \& $r'$ bands we ignored the
four faintest points in each of these bands. The amplitudes derived
when we included these points in the analysis are given in parenthesis.

It appears as if the amplitude of the modulation decreases with
decreasing wavelength, from $i'$ to $B$, even when ignoring the four
faintest points in $i'$ \& $r'$. The origin of this varying modulation
is not clear. It could result from some real wavelength dependence of the
system's modulation. On the other hand, the
different modulation could
come either from some secular changes during the $\sim~\!\!\!\!4$ years
between the KPNO $B$ \& $V$ and CFHT $i'$ \& $r'$ observations, or
from some wavelength-dependent blending of light from a near-by star.

\begin{table}
  \caption{Harmonic fits amplitude and minimum phase near 0.0.}   
  \label{harmonicfit}  
  \centering         
  \begin{tabular}{c l l}
    \hline\hline
    Filter & \, Amp.                    & \,\, Min. Phase   \\
           & [mmag]                  &                   
\\
    \hline
$i'$& $47 \pm 3 \,\, (63 \pm 4)$ & $0.962 \pm 0.006 \,\, (0.949 \pm 0.006) $ \\
$r'$& $39 \pm 2 \,\, (48 \pm 2)$ & $0.960 \pm 0.003 \,\, (0.953 \pm 0.003) $ \\
    $V$    & $34 \pm 1$                        & $0.970 \pm 0.001$ \\
    $B$    & $30 \pm 1$                        & $0.979 \pm 0.001$ \\
    \hline
  \end{tabular}
\end{table}

\section{Modelling the system}

In our modelling of the X-7 binary system, we preferred to find first a
stellar model that fits the broad-band photometry of the system,
assuming the optical brightness is coming from the optical star
alone. From the stellar radius, the observed X-ray eclipse width and
the periodic ellipsoidal modulation we can derive the two masses as a function
of the orbital inclination.

\subsection{Optical star modelling}
\label{photometricmodelling}

We have used the following broad-band magnitudes: $U$ =
18.1, $B$ = 18.8, $V$ = 18.9 \citep[][ based on HST data]{Pietsch06},
$R$ = 19.0 \citep[][ LGGS\footnote{The Local Group Galaxy Survey,
http://www.lowell.edu/users/massey/lgsurvey/index.html}]{Massey06},
$r'$ = 19.1 and $i'$ = 19.42 mag \citep{Hartman06}, assuming an
uncertainty of 0.1 mag on all values. We converted the observed
measurements to absolute magnitudes by using an M33 distance modulus
of 24.62 mag \citep{Freedman01}, $A_V = 0.53 \pm 0.06$ mag from
\cite{Pietsch06} and $A_{\lambda}/A_V$ values from \cite{Schlegel98}.

We compared these magnitudes to the ones determined from the
integrated magnitudes from \cite{Girardi02, Girardi04}, where we
assume $M_{bol,\sun} = 4.77$ \citep{Girardi02} and fit for the
radius for each value of $\log(g)$ and $T_{eff}$.  We did this for
both [Fe/H] = 0 and [Fe/H] = -0.5 tables, since assuming X-7 is at a
galactocentric distance of 2 kpc results in [Fe/H] = -0.22, based on
the [O/H] gradient of \cite{Garnett97} and the [O/H] to [Fe/H]
conversion of \cite{Maciel03}.

\begin{figure}
  \centering
  \includegraphics[width = 6cm]{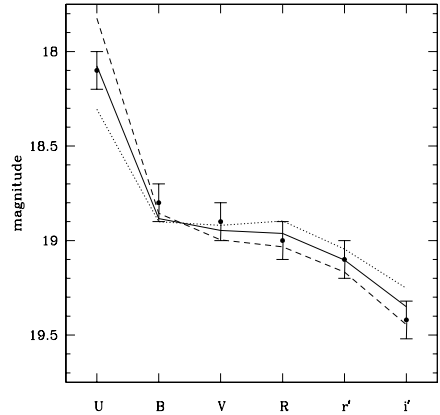}
  \caption{Photometric broad-band modelling of the X-7 stellar
     component.  Error bars present the assumed uncertainty of 0.1 mag
     in all bands. Solid line presents a model with $T_{eff}$ = 27000
     K, $\log g = 5.0$ and $R = 25\,R_{\sun}$. The dashed and dotted
     lines the models at 95\% confidence boundary of our
     modelling. The dotted line is a model with $T_{eff}$ = 18000 K,
     $R = 35\,R_{\sun}$, the dashed line is a model with $T_{eff}$ =
     47000 K, $R = 15\,R_{\sun}$.}
  \label{phot}
\end{figure}

The data allow a large range of stellar models. One such model, for
[Fe/H] = 0, with $\chi^2 = 1.62$, has $T_{eff} = 26000$ K, $\log g =
5.0$ and $R = 25\,R_{\sun}$. For [Fe/H] = -0.5, with $\chi^2 = 1.55$,
we get $T_{eff} = 27000$ K, $\log g = 5.0$ and $R = 25\,R_{\sun}$. The
95\% uncertainty range, where $\chi^2$ increases by up to 8.0 (for
three parameters), is: $18000 < T_{eff} < 47000 $ K, $15 < R <
35\,R_{\sun}$. The value of $\log g$ is unconstrained. Fig.~\ref{phot}
brings three representative models; the solid line presents the model
with $T_{eff} = 27000 $ K, $R = 25\,R_{\sun}$, the dotted line
represent the model with $T_{eff} = 18000 $ K, $R = 35\,R_{\sun}$ and
the dashed line shows the model with $T_{eff} = 47000 $ K, $R =
15\,R_{\sun}$.  The figure shows that the absolute magnitudes of the
X-7 optical counterpart can not constrain the stellar radius, and any
value in the range of 15--35 $R_{\sun}$ is possible. However, as we
show in the next section, a stellar radius of $R > 20\,R_{\sun}$ is
very unlikely, since the star will be too massive. Therefore, our most
likely models are in the range of $15 < R < 20 \,R_{\sun}$ and $33000
< T_{eff} < 47000 $~K.

\subsection{Light-curve modelling}
\label{lightcurvemodelling}

We used two different codes, PHOEBE and ELC, to model the obtained
periodic light curves and derive an estimate of the masses of both
components.  For both codes, the main effect is due to the
tidally induced ellipsoidal shape of the optical component.  

The first program is the "PHysics Of Eclipsing BinariEs"
program of \cite{Prsa05}, a front-end code for the Wilson-Devinney
program \citep{Wilson71, Wilson79, Wilson90}. Following
\cite{Pietsch06}, who used PHOEBE to model the X-7 $B$ and $V$ light
curves, we used the eclipse half angle value of $26.5 \degr \pm 1.1
\degr$ to find the ratio of the radius of the optical component to the
semi-major axis as a function of inclination.  We varied the
inclination between $90 \degr$ and about $70 \degr$ and fit for the
mass ratio and a luminosity scale to match each light curve, assuming
a temperature of 27000 K and linear limb-darkening
coefficients from \cite{Claret00, Claret04}, assuming a metallicity of
[Fe/H] = -0.2.

We performed the fits ignoring the four faintest points in each of the
$r'$ and $i'$ light curves which occur near phases 0 and 0.5 --- no
model was able to match these points. Including those points does not
substantially change the mass-ratio, while it does slightly affect the
luminosity scale.

\begin{figure*}
  \centering
  \includegraphics[width = 12cm]{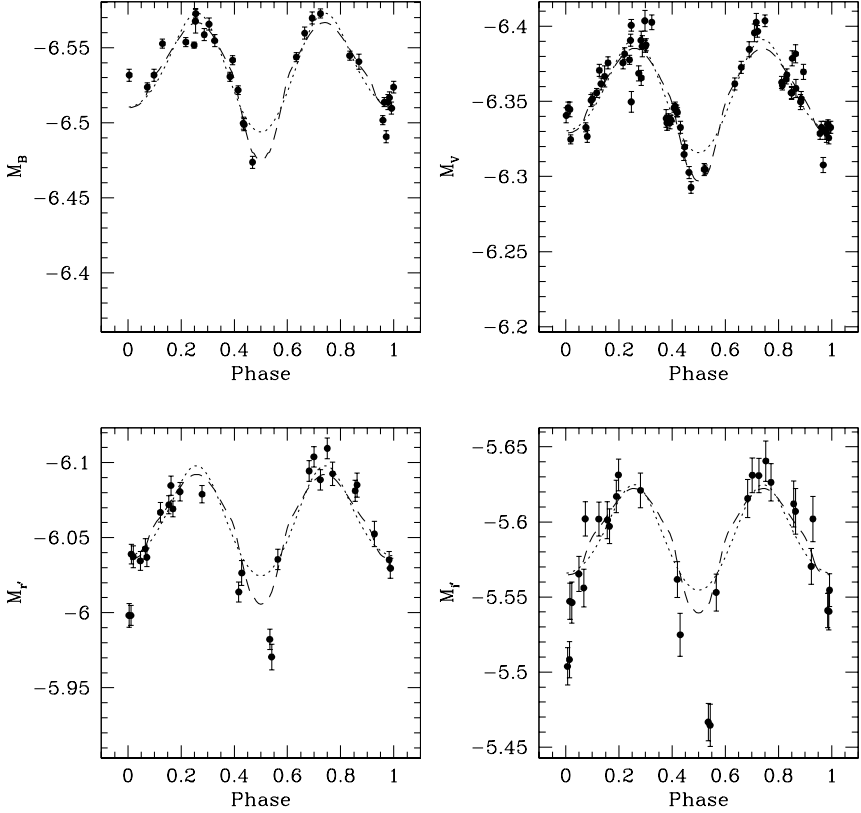}
  \caption{$BVr'i'$ light curves phased at $P$ = 3.453014 day and
    $T_0$ = 2453639.119 from \cite{Pietsch06}, with model light curves 
    for $R=20\,R_{\sun}$, 
    $T_{eff}$ = 33000 K and $i = 80 \degr$, overplotted. The dotted line 
    is the PHOEBE light curve, with no accretion disk or X-ray
    heating. The dashed line is the ELC model which includes an accretion
    disk and X-ray heating.
    }
  \label{lcmodel}
\end{figure*}

There is not enough data to constrain the stellar radius or the
inclination of the system; a good fit to the light curves could be
found for large ranges of these two parameters.  One such
model is plotted in Fig.~\ref{lcmodel}. The PHOEBE light curve is
drawn with a dotted line and the other model with a dashed line. For
both, values of $R = 20\,R_{\sun}$, $T_{eff}$ = 33000 K and $i = 80 
\degr$ were taken, although changing these values results in
essentially no difference for the model light curve.

The range of allowed inclinations and stellar radii did not enable us
to constrain the masses of the two components and they could only be
derived as a function of the assumed orbital inclination and optical
radius.  We found that for the PHOEBE code the mass ratio
($M_x/M_{opt}$) varies from 0.29 at $90 \degr$ to 0.15 at $70
\degr$. Below $70 \degr$ the optical component would exceed its Roche
lobe. Assuming $R = 20\,R_{\sun}$, the mass of the optical component
would vary between $79\,M_{\sun}$ and $49\,M_{\sun}$, while that of
the compact object would vary between $23\,M_{\sun}$ and
$7.5\,M_{\sun}$. Because the fits are fairly insensitive to the
temperature of the optical component, the resulting masses scale as
the radius of the star to the third power. For $R = 15\,R_{\sun}$, the
mass of the optical component varies between $33\,M_{\sun}$ and
$21\,M_{\sun}$, while the compact object mass varies between
$10\,M_{\sun}$ and $3\,M_{\sun}$. We note, however, that assuming a
stellar radius larger than $20\,R_{\sun}$ results in an unrealistic
large stellar mass. We therefore limit the discussion to stellar radii
smaller than $20\,R_{\sun}$.

The second program used was the ELC code of \cite{Orosz00}.  This
program has been specifically tailored to model optical light curves
of X-ray binary systems, including effects such as the presence of an
accretion disk, X-ray heating of the optical component and the ability
to explicitly take one of the components to be invisible. To model the
X-ray heating we assumed an average X-ray luminosity of $10^{37.5}$
erg s$^{-1}$, and to model the disk we assumed an inner disk
temperature of $10^7$ K \citep{Pietsch06}.  We assumed a disk opening
angle of $2 \degr$, and a power-law exponent for the temperature
profile of the disk of $-0.75$. We used the black-body mode of the
model rather than using the detailed model atmosphere since the model
atmosphere would not necessarily be a good match to the Sloan $r'$ or
$i'$ filters. We proceeded as above, stepping through a range of
inclination angles for two values of the optical radius and fitting
for the mass ratio. This time we also fit for the inner and outer disk
radii. One possible model, for $R=20\,R_{\sun}$, $T_{eff}$ = 33000 K
and $i = 80 \degr$, is plotted in Fig.~\ref{lcmodel}.

The resulting mass ratio varies from 0.23 at $90 \degr$ inclination to
0.07 at $65\degr$ inclination. For a radius of 20 $R_{\sun}$ this
corresponds to a mass range of 83 to 42 $M_{\sun}$ for the optical
component and a range of 19 to 3 $M_{\sun}$ for the compact
object. Below $65 \degr$ the ELC model star would exceed its Roche
lobe. In units of the inner Lagrange radius of the compact object, the
inner disk radius varies from $1.6 \times 10^{-5}$ to $5.4 \times
10^{-5}$ while the outer disk radius varies from 0.61 to
0.81.

\begin{figure*}
  \centering
  \includegraphics[width = 10cm]{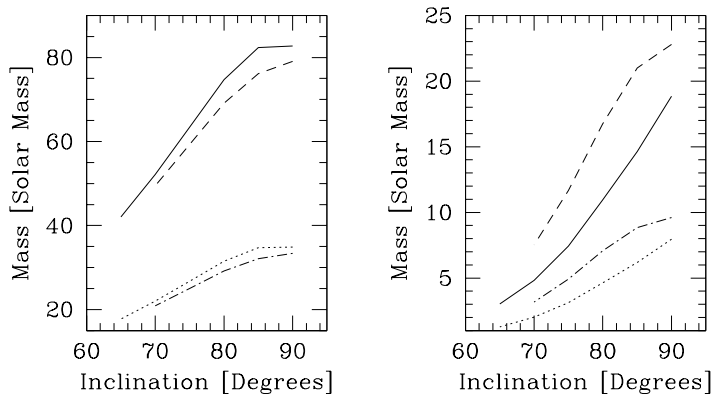}
  \caption{Plots showing the mass vs. inclination for two values of
  the stellar radius. The left plot is for the optical component, the
  right plot is for the compact object.  The solid and the dashed
  lines show the results for $R = 20\,R_{\sun}$, using the ELC and the
  PHOEBE codes, respectively. The dotted and the dot-dashed lines show
  the results for $R = 15\,R_{\sun}$, using the ELC and the PHOEBE
  codes, respectively.}
  \label{mvi}
\end{figure*}

The masses of both components as a function of inclination, derived by
both codes for two values of the optical radius, are plotted in
Fig.~\ref{mvi}.

\section{Discussion}
\label{summary}

We have conducted a broad-band optical analysis of the X-7 optical
counterpart, including period analysis, photometric modelling and light
curve modelling. We applied a multi-band double-harmonic period
analysis of X-7 optical light curves in four bands and identified a
period of $P_{opt}\!=$ \period. This period, derived from an
independent optical analysis, is in good agreement with the known
X-rays period \citep{Pietsch06}.

Using photometric absolute magnitudes in six bands, we find a range of
models for the optical counterpart, with $15 < R < 35 \,R_{\sun}$ and
$18000 < T_{eff} < 47000 $ K. However, stellar radius larger than
$20\,R_{\sun}$ results in an unrealistic large stellar mass. We
therefore suggest that the stellar radius is smaller than
$20\,R_{\sun}$. The possible temperature range is reduced accordingly
to $33000 < T_{eff} < 47000$.  The classification of the optical star
as O6 III by \citet{Pietsch06} is well within our uncertainties.

We have modelled the optical light curve using two programs. While the
ELC model, incorporating both a disk and X-ray heating matches better
to the data than the PHOEBE model (see Fig.~\ref{lcmodel}), neither
model yielded a good fit to the light curves. In particular, the
minima in $r'$ and $i'$ appear to be underestimated by all models that
we have tried. At this point it is unclear if this is due to random or
systematic errors in the observations or to inadequacies in the
physical models.

The simplistic models can nevertheless yield a rough estimate of the
masses of the two components as a function of the optical radius and
the orbital inclination. For the smallest likely optical radius 
($15\,R_{\sun}$) we get $3\,M_{\sun}$ and $1.3\,M_{\sun}$ from the 
PHOEBE and ELC codes respectively. Therefore, the present analysis can
not exclude completely a neutron star as the compact object, although
this option is very unlikely. 

On the other hand, the similarity between M33 X-7 and the three known
black-hole HMXB --- Cyg X-1, LMC X-1 and LMC X-3
\citep[e.g.,][]{Cowley92,mcClintock03}, is striking. All four systems
have orbital periods of a few days, their optical counterpart is an
early-type star and the compact object has a mass of 6--10
$M_{\sun}$. Radial velocity measurements of M33 X-7 will enable us to
better constrain the masses of the two components.  Large telescopes
and presently efficient spectrographs render such measurements
feasible. Together with the X-ray eclipse and the optical light
curves, we should be able to understand this system better than any
other black-hole HMXB.

Finally, if this system would be proven to be a black-hole HMXB, it is
interesting to note that three out of the four such known systems were
found in external galaxies. This might indicate that we are not very
efficient in detecting such systems in our own galaxy. Maybe the dust
in the Galactic plane hides from our telescopes many more Galactic
black holes.

\begin{acknowledgements}

We are grateful to J. McClintock for helpful discussions and to J. 
Orosz for sharing his ELC program with us.
We thank the anonymous referee for his helpful comments. 
This work was partially funded by the German-Israeli Foundation for
Scientific Research and Development and by the Israeli Science
Foundation. This research has made use of NASA's Astrophysics Data
System Abstract Service and of the SIMBAD database, operated at CDS,
Strasbourg, France.

\end{acknowledgements}

\bibliographystyle{aa}
\bibliography{0030bib}

\end{document}